\documentclass[twocolumn,twoside,slac,floatfix]{revtex4}
\usepackage{graphicx}
\usepackage{fancyhdr}
\begin{document}

\title{MAGIC: Exact Bayesian Covariance Estimation and Signal Reconstruction
       for Gaussian Random Fields
      }


\author{Benjamin D.~Wandelt\footnote{NCSA Faculty Fellow}}
\affiliation{Department of Physics, University of Illinois at Urbana-Champaign, IL 61801, USA}
\affiliation{Department of Astronomy, University of Illinois at Urbana-Champaign, IL 61801, USA}

\begin{abstract}
In this talk I describe MAGIC \cite{MAGIC}, an efficient approach to covariance
estimation and signal reconstruction for Gaussian random fields (MAGIC Allows
Global Inference of Covariance). It
solves a long-standing problem in the field of cosmic
microwave background (CMB) data analysis but is in fact a general technique
that can be applied to noisy, contaminated and incomplete or censored
measurements of either spatial or temporal Gaussian random
fields. In this talk I will phrase the method in a way that emphasizes its general structure and applicability but I comment on applications in the CMB context.  The method allows the exploration of
the full non-Gaussian joint posterior density of the signal and
parameters in the covariance matrix (such as the power
spectrum) given the data. It generalizes the familiar Wiener filter in
that it automatically discovers signal correlations in the data as
long as a noise model is specified and priors encode what is known
about potential contaminants. The key methodological difference is that
instead of attempting to evaluate the likelihood (or posterior
density) or its derivatives, this method generates an asymptotically exact Monte Carlo sample from
it. I present example applications to power
spectrum estimation and signal reconstruction from measurements of the
CMB. For these applications the method achieves speed-ups of
  many orders of magnitude compared to likelihood maximization techniques, while offering greater flexibility in modeling and a full characterization of the uncertainty in the estimates.\end{abstract}

\maketitle

\thispagestyle{fancy}


\section{Introduction}
Signal reconstruction from noisy data is one of the {\em raisons d'\^etre}
of applied  statistics. If the signal is a
Gaussian random field, and the signal and noise covariances are known
in advance, Wiener filtering \cite{wiener,rybickipress} is the theoretically
optimal method for estimating the signal from noisy data. In this simple case the
solution is a linear operator that acts on the data vector and
returns the minimum variance, maximum likelihood and maximum a
posteriori estimator of the signal given the data.

What ought to be done, however, if the signal covariance is not known
in advance, and the signal covariance must  be estimated from the
data? In fact there are applications where covariance estimation is
the primary goal and signal reconstruction is secondary. These cases
have traditionally been treated separately. For stationary
signals, the covariance of the signal is best specified in the Fourier
basis since this basis diagonalizes the covariance matrix. In these
cases covariance estimation becomes power spectrum estimation.
One such example is  cosmic microwave background data (CMB) analysis which motivated this
analysis. I will return to it in section \ref{CMB}. Other examples are time series analysis, spatial analysis of censored data, such as geological surveys, power spectrum estimation and signal reconstruction for helioseismology, image reconstruction based on a stochastic model of the form of pixel-pixel correlations, etc. The method described here generalizes the results of \cite{rybickipress} and should therefore also be useful for the applications discussed there.

In this talk I will first review the common structure that underlies
these apparently different statistical problems (section \ref{review}).
I will then summarize the main advances realized by the new method in
section \ref{method}. The subsequent section contains the results
from the application of this new approach to the first all-sky CMB data set. Further
details and examples can be found in our paper \cite{MAGIC} and online materials at the conference WWW site. \footnote{PowerPoint slides of
  this talk and two AVI movies  are at
 http:// www-conf.slac.stanford.edu/phy-stat2003/talks/wandelt/contributed/}

The ideas in this paper were developed from a Bayesian perspective.
There are pros and cons of Bayesian estimation. The pros are many: it maximizes the use of all available information and treats measurements, constraints and model on the same footing as information.  The result of a Bayesian estimation is a probability density, not just a number, so one automatically obtains uncertainty information about the estimate. However, if Bayesian methods are implemented naively, these advantages come at the price of heavy computation especially for  multivariate problems. However the results presented in this paper are an example that it is possible to overcome these computational challenges and make Bayesian techniques work in a highly multivariate ($D\sim 10^6$) problem.

\section{Signal Reconstruction and
Covariance Estimation}
\label{review}
In this section I will review the problems of signal reconstruction and
covariance estimation from a Bayesian perspective. First, some notation.  Let us
assume that the data were  taken according to the model equation
\begin{equation}
d=A(s+f) +n
\label{model}
\end{equation}
where the $n_d$-vector $d$ contains the data samples, the ($n_d\times
n_s$) matrix
$A$ is the observation matrix, the $n_s$-vector $s$ is the
(discretized) signal, the $n_s$-vector $f$ represents any
contaminants (``foregrounds'') one has to contend with, and the  $n_d$-vector $n$ is the instrumental noise. I model the signal stochastically (vs. a deterministic functional form) and ``infer'' its covariance properties from the data. In particular, the signal is modeled through its covariance properties, encoded in $S\equiv\langle s s^T\rangle$, the  signal covariance matrix.

Then I can write the Bayesian posterior as
\begin{equation}
P(s,f,S|d,N)\propto P(d|s,f,N)P(s|S)P(f)P(S)
\label{bayes}
\end{equation}
where $N$ is the noise
covariance matrix $\langle n n^T\rangle$. I will now discuss the various terms in
Eq.~\ref{bayes}.
The likelihood $P(d|s,f,N)$ specifies how the data is related to the quantities
in the model. Given the model equation Eq.~\ref{model} specifies that
$P(d|s,f,N)=G(d-A(s+f) ,N)$\footnote{I use $G(x,X)$ as a shorthand for the
multivariate Gaussian density
\begin{equation}G(x,X)=\frac{1}{\sqrt{|{2\pi X}|}}\exp\left({-\frac12 x^T X^{-1} x}\right).
\end{equation}
}.

The other terms in Eq.~\ref{bayes} specify information about the components of
the model. The term $P(s|S)$ contains information about the covariance of $s$.
If $s$ is a  Gaussian random field with zero mean (examples from cosmology are
the CMB or other probes of the density fluctuations of
matter on cosmological scales) $P(s|S)=G(s,S)$. Note that it is not assumed that
$S$ is known.

Partial knowledge (or ignorance) about $S$ is quantified in
terms of the prior $P(S)$. For a stationary field $P(S)$ might simply represent
the fact that I parameterize the covariance matrix in terms of power
spectrum coefficients. Eq.~\ref{bayes} also assumes that the signal, noise and the
contaminants are stochastically independent of each other. Further, the equations
as written are conditioned on perfect knowledge of the noise
covariance.\footnote{This assumption may not hold in practice and can in fact be
relaxed. The resulting question whether both $S$ and $N$ can be usefully obtained
from the data is determined by the structure of the observation matrix $A$.}

Lastly, $P(f)$ encodes the knowledge or ignorance about foregrounds. Note that
from a Bayesian perspective all that is required is that $P(f)$ accurately
represents knowledge about $f$. Therefore assuming a Gaussian
form for $f$ does not assume that $f$ actually has Gaussian statistics.
In particular the mode of the Gaussian corresponds to the  most probable (a priori) foreground model and the covariance to
the uncertainty in the model. The ability to specify uncertainties in
the foregrounds (which will then be taken into account when the method is
applied) is a key feature of this approach which guards against biases from
including incorrect foreground templates without the ability to account for
the uncertainty in these templates.

Having specified the forms of the various
terms on the right hand side of Eq.~\ref{bayes}, the task is to explore the
joint posterior density $P(s,f,S|d,N)$. However, traditionally the problem
is treated in three different limits. If, as an expression of prior ignorance, I take
$P(f)=const.$ and $P(s)= \int P(s|S)P(S)dS=const.$ then all
the information is in the likelihood $P(d|s,f,N)$. In this case the best
one can do if $n$ is Gaussian, is to summarize what is known about $s+f$ in terms
of the maximum likelihood estimate
\begin{equation}
m\equiv(A^TN^{-1}A)^{-1}A^TN^{-1}d
\end{equation}
and quote the associated noise covariance matrix $C_N=<m
m^T>=(A^TN^{-1}A)^{-1}$. In the CMB literature the  process of obtaining $m$ and
$C_N$ from the data are known as ``map making.''

If on the other hand, the signal
covariance $S$ is perfectly known and foregrounds are neglected then the joint
posterior becomes $P(s,S|d,N)\propto P(s|d,N,S)$ where
\begin{equation}
P(s|d,N,S)=G(s-S(S+C_N)^{-1}m,S(S+C_N)^{-1}C_N).
\label{Pofsgivend}
\end{equation}
This posterior for $s$ peaks at $s_{WF}$, the well-known Wiener Filter
reconstruction of $s$, so this is known as ``Wiener Filtering.''

In the third limit, ``power spectrum
estimation,'' one does not know $S$ but have some information about how it
is parameterized, namely that in the Fourier basis $S$ is diagonal with
the diagonal elements equal to the power spectrum coefficients $C_l$. If we
ignore foregrounds again and set $P(S(C_l))=const$  we can integrate
out (``marginalize over'')
 $s$ and obtain the usual starting point for maximum likelihood power
spectrum estimation
\begin{equation}
P(S(C_l)|d,N)=G(m,S(C_l)+C_N).
\label{PofSgivend}
\end{equation}

The density $P(S(C_l)|d,N)$ is considered as a multivariate function of all the
power spectrum coefficients up to some band limit $l_{max}$. It represents all the
information about $S(C_l)$ contained in the data. One can again
summarize what is known about $S$ by quoting the set of power spectrum estimates
$\hat{C}_l$ for which $P(S|d,N)$ is maximum (equivalent to the maximum likelihood
estimates) and include a summary of the width of the
marginal distribution of $P(S|d,N)$ for each power spectrum coefficient.

However, in this case for any $n_s$ larger than a few thousand this
procedure is computationally prohibitive. Since the determinant in Eq.~\ref{PofSgivend}
depends on $S$, it needs to be evaluated if the shape of the likelihood is
to be explored. Determinant evaluation scales as $n_s^3$. As a result,
to evaluate
Eq.~\ref{PofSgivend} just once for a
million pixel map would take several
years, even if one achieved perfect
parallelization across thousands of processors on the
most powerful supercomputing platforms in the world.  To find
its maximum in a parameterization of 1000 power spectrum coefficients
and compute marginalized confidence
intervals for each $C_l$ by integrating out all others is a lost cause.

The maximum likelihood techniques that are currently described in the
literature \cite{Tegmark,BJKcl} avoid the determinant calculation in
Eq.~\ref{PofSgivend} by finding the zero of the first
derivative of $P(S|d,N)$ using an approximate Newton-Raphson iteration
scheme. However, for realistic data, the computational complexity is not
reduced because the first derivative contains traces of matrix
products that also require of order $n_s^3$ operations. In these
treatments the error bars on the power spectrum coefficients are
approximated by the second derivative of the likelihood at the
peak even though the likelihood of $S$ is non-Gaussian. This second
derivative is again hard to compute, requiring of order $n_s^3$ operations.

Even these expensive methods do not provide a way of accurately summarizing and publishing the ``data product,'' $P(S|d,N)$. There are various approximate techniques for doing this in the literature \cite{BJK,WMAPverde} but it is not well understood how good these approximations are away from the peak of the likelihood \cite{lewisproc}.

\section{Method: Do Not Evaluate, Sample!}
\label{method}
How does one overcome these computational challenges? The answer I propose is to {\em sample}
from the full joint density $P(s,f,S|d,N)$. This may seem even more challenging,
since this a function of millions arguments and general techniques of generating
samples from complicated multivariate densities are very computationally
intensive. However, the special structure of the Gaussian priors in Eq.
\ref{bayes} allows exact sampling from the conditional densities of
$P(s,f,S|d,N)$. Exact sampling is made possible by solving systems of equations
using the preconditioned conjugate gradient method \cite{NR}. This means the
{\em Gibbs sampler} \cite{Tanner} can be used to construct a Markov Chain which will
converge to sampling from $P(s,f,S|d,N)$. The Gibbs sampler is an iterative scheme
for generating samples from a joint posterior density by iterating over the
components of the density (such as $s$, $S$, and $f$) and sampling each of them in turn from their conditional distributions while
keeping the other components fixed. Given a set of Monte Carlo samples from the
joint posterior, any desired feature of the posterior density can be computed with accuracy only limited by the sample size.

After having obtained a sample from the joint posterior $P(s,f,S|d,N)$, it is trivial to generate samples from the marginal posteriors $P(s|d,N)$ or $P(S|d,N)$. Integration over a sampled representation of a function just corresponds to ignoring the dimensions that are being integrated over! For the problem at hand things are even better than this, since the conditional density $P(S|s)$ has a very simple analytical form. As a result, one can compute an analytical approximation to $P(S|d,N)$ using the Monte Carlo samples $s_i$
\begin{eqnarray}
P(S|d,N)=\int ds P(S|s) P(s|d,N)\approx \nonumber \\
(1/n_{MC}) \sum_{i=1}^{i=n_{MC}} P(S|s_i).
\label{margpost}
\end{eqnarray}
This is known as the Blackwell-Rao estimator of $P(S|d,N)$ which is guaranteed to have lower variance than a binned estimator. In fact one can show that for perfect data (complete and without noise) this approximation is exact for a Monte Carlo sample of size 1! For realistic data, the approximation
converges to the true power spectrum posterior given enough samples.

My collaborators and I call the approach and the set of tools we have developed to implement this approach the ``MAGIC'' method, since MAGIC Allows Global Inference from Correlated data. We give a detailed description of the technique in the context of CMB covariance analysis in \cite{MAGIC}.
Figure~\ref{performance} shows the performance of MAGIC compared to power
spectrum estimation techniques (which do not include the signal reconstruction
and foreground separation features of MAGIC).The main advantages of the MAGIC method are the
following:
\begin{figure}
\centering
\includegraphics[width=80mm]{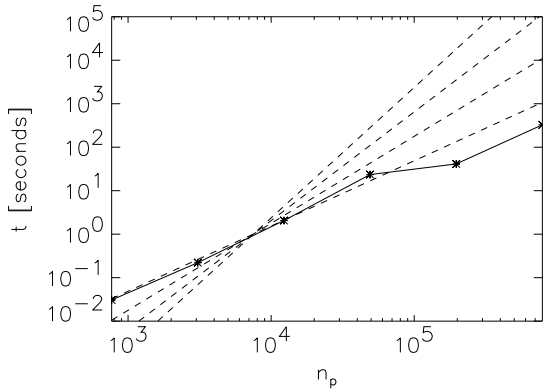}
\caption{Average computing time (without code optimization) required for one iteration of the Gibbs sampler
as a function of the number of pixels in the  map. These timings are for a
single AthlonXP 1800+ CPU. Solid line:  actual timings. Dashed lines show
$n_p^x$ for  $x\in\{3,5/2,2,3/2\}$ from the top to the bottom on the right side
  of the figure. Brute force methods require $t\sim O(n_p^3)$  and approximate
methods require $t\sim O(n_p^{(3/2)})$ computational time. For the WMAP data $n_p \sim 3\times 10^6 $ pixels.} \label{performance}
\end{figure}\begin{enumerate}
\item Massive speed-up compared to brute force
methods. For an (unrealistic) pre-factor of 1 a single $n_p^3$ operation would take $3\times 10^{10}$ seconds on a 1 GFlop computer. An unoptimized implementation running in the background on a desktop AthlonXP1800+ CPU currently requires less than $10^5$ seconds per sample.
\item Massive reduction in memory use: since we only need to compute matrix-vector products (not matrix-matrix products, matrix inverses or determinants) only the parametrizations of the covariance matrices need to be stored (e.g. noise power spectrum for $N$ and the signal power spectrum for $S$). This reduces the memory requirements from order $n_p\times n_p$ to at most order $n_d$ which is usually many orders of magnitude less.
\item Allows
modeling realistic observational strategies and instruments.
\item Straightforward
parallelization (run several MAGIC codes on separate processors to generate
several  times the number of samples in the same time).
\item Allows treating the
statistical inference problem globally, that is it keeps the full set of
statistical dependencies in the joint posterior given the data.
\item Generalizes Wiener Filter signal reconstruction to situations where the signal
covariance is not known a priori but automatically discovered from the data at
the same time as the actual signal is reconstructed.
\item Allows computing
marginal credible intervals, either for individual power spectrum
estimates or for combinations of any set of dimensions in the very high
dimensional parameter space.
\item Allows incorporating uncertainties (e.g.
about the foregrounds) in the analysis in such a way that they are propagated
correctly through to the results.
\item Makes it possible to build in physical
constraints in a straightforward way.
\item Generates an unbiased functional
approximation to $P(C_l|d)$, as shown in Eq.~6. It has the advantage of being a controlled and improvable approximation and removes
the need for parametric fitting functions such as the offset log-normal or
hybrid approximations.
\item Generates a {\em sampled} representation of the joint posterior Eq.~\ref{bayes}, which
simplifies further statistical
analyses.\end{enumerate}

Since MAGIC is a Markov Chain method, one also has to discuss the issue of burn-in and correlations of subsequent steps in the chain. Steps in the power spectrum coefficients $C_l$ are proportional to the width of the perfect data posterior \cite{MAGIC}. In other words, the number of steps it takes to generate two uncorrelated power spectrum samples is proportional to $(S/N)_l^2$ where $(S/N)_l$ is the rms signal to noise ratio for the $l^{\rm th}$ power spectrum coefficient. Conveniently, the samples are nearly uncorrelated over the range in $l$ where the data is informative. In numerical experiments with the WMAP data it took about 15-20 steps for the chain to burn-in (for the range in $l$ where $(S/N)_l\sim 1$ or greater) from a wildly wrong initial guess of the power spectrum ($C_l=const.$).

\begin{figure}
\centering\includegraphics[width=62mm]{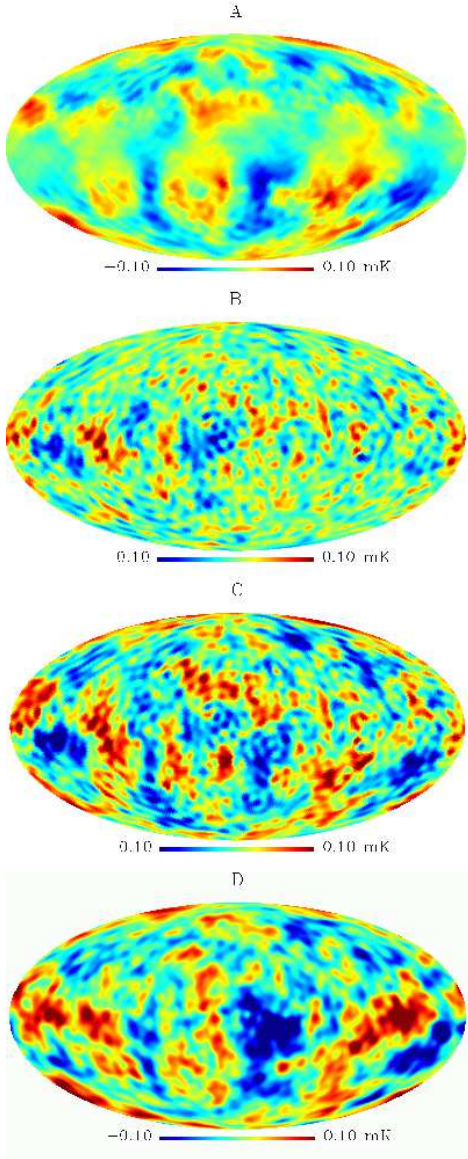}
\caption{Reconstructed signal maps in Galactic coordinates. A: The posterior mean signal map ($\int ds s P(s|d))$ for the COBE-DMR data. This is a  generalized Wiener filter which
does not require knowing the signal  covariance a priori. B: One sample drawn
from the  conditional posterior $P(s|C_l,f,d)$. The posterior mean signal map, shown in panel A, has been removed. C: The  sample pure signal sky at
the same iteration. This is the pixel-by-pixel sum of the maps in panels A and B. D: The WMAP data smoothed to 5 degrees (less than A, more than C). The corresponding
features in parts A and D are clearly visible.}
\label{maps}
\end{figure}

\section{Example Applications to the Cosmic Microwave
Background}\label{CMB}
In the online materials for this talk (see footnote 1) I present
the results of applying the MAGIC method to a synthetic data set which covers an
unsymmetrically shaped part on the celestial sphere. I used MAGIC to reconstruct the
signal on the full sky and to make movies of the Gibbs sampler iterations. This
is an example where the signal is automatically discovered in the data by
the algorithm, without specification of the signal covariance.

Figures~2 and 3 show the results of analyzing the COBE-DMR data \cite{COBE},  one of the
most analyzed astronomical data sets. This allowed us to perform
consistency checks between the MAGIC method, other methods and the  recent results
from the Wilkinson Microwave Anisotropy Probe (WMAP)
\cite{WMAP}.

\begin{figure}
\centering\includegraphics[width=80mm]{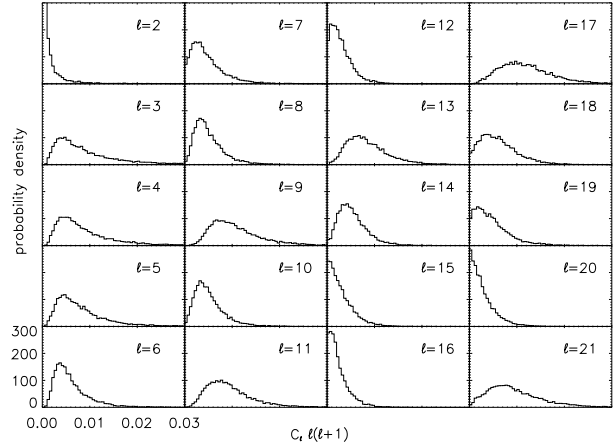}
\caption{Marginal posterior densities for each individual $C_\ell$  from
the COBE-DMR data. At each $\ell $ the fluctuations in the $C_\ell$ at  all
other $\ell $ were integrated out. The axis ranges are the same  for all
panels.} \label{COBEcls}
\end{figure}

I am also very interested in evaluating
claims that the WMAP data favors theories which predict a lack of large scale
fluctuation power in the CMB. This claim, if true, would have far-reaching
consequences for our understanding of the Universe. Since cosmologists only have one sky
to study, we have to be very careful to account for our limited ability
to know the ensemble averaged power spectrum on large scales. The WMAP team
estimated the fluctuation power on large scales using several techniques and
consistently found it to be low. However, in all of these techniques, the
variance of the estimates was computed in an approximate way (e.g. in terms of
the curvature at the peak) and relies on theory for the assessment of
statistical significance.
Using MAGIC one can easily integrate over the
posterior density of the power spectrum given the data. Therefore it is easy to compute
the probability for the power spectrum coefficients in any given $l$-range
to be smaller than any given value.

Using the MAGIC method it was straightforward to generate a preliminary sample of the
power spectrum coefficients from the WMAP posterior using only the W1 channel,
one of the cleanest channels in the WMAP data, in terms of systematic error
estimates. For the cleaned W1 data and masking regions of galactic emission (mask {\it Kp0} in the WMAP data release) the quadrupole and octopole power is not obviously discrepant from theoretical expectations. Choosing a more aggressive mask could change this since that reduces the sampling variance. One should bear in
mind that the power spectrum likelihood $P(C_l|d)$ has  infinite
variance for $l=2$ even for perfect all-sky data, unless a prior is put on $C_2$'s
value. Therefore, in an exact assessment of the quadrupole issue claims of a significant
discrepancy ought to be based on the actual shapes of posterior density, not a chi-square test
(compare the detailed discussion of cosmic variance in \cite{MAGIC}).
I will
address the issue of low power in the low cosmological multipoles in a future
publication.

\section{Future Directions}

%The analysis so far has been conditioned on the assumption that the signal $s$ is Gaussian. For CMB analysis it would be a physically very important discovery if a non-Gaussian stochastic model provided a better fit. While it is easy to write down physically meaningful non-Gaussian models, implementing the analysis raises practical difficulties, since sampling from the conditional densities may be complicated. Traditionally, tests for non-Gaussianity have been implemented in a frequentist way: define a statistic, compute it on the data and on Monte Carlo simulations which were constructed under the null hypothesis. Then assess consistency.

%Using the MAGIC framework, one could generate the posterior density of the non-Gaussianity statistic from the posterior density of the signal $s$ and then test whether it is significant. Since a Gaussian prior was used to construct the posterior density of $s$ I would expect the non-Gaussianity in $s$ to be reduced, but if the data contained enough evidence for non-Gaussianity it would prevail.

Of course, if desired, additional prior information about our Universe can
be added to the analysis. For example instead of viewing the power spectrum as the quantity of interest, its shape could be parameterized as a function of the $\sim 10$  cosmological parameters which span
the space of cosmological theories. Then instead of sampling from the power
spectrum coefficients given the signal, one would run a short Metropolis-Hastings Markov chain at each Gibbs iteration to
obtain a sample from the space of cosmological parameters given the data. These parameter samples, in turn define a density over the space of power spectra with considerably tighter error bars.
The result is the non-linearly optimal filter for reconstructing the mean of the power spectrum incorporating
physical information about the origin of the CMB anisotropies.

Another important direction is the analysis of image distortions. The treatment as
detailed so far does not allow for the CMB to be lensed gravitationally
by the mass distribution through which it streams on its way to us. This
distortion itself contains very valuable cosmological information. Extending the formalism to
account for lensing of the CMB and estimate the statistical
properties of the lensing masses from the lensed CMB would be an important extension of this
approach.

\begin{acknowledgments}
I thank my students and collaborators Arun Lakshminarayanan, David Larson, and Ian O'Dwyer, as well as Tom Loredo for his suggestions. This work has been partially supported by the
National Computational Science Alliance under grant number
AST020003N and the University of Illinois at Urbana-Champaign.

\end{acknowledgments}

\end{document}